\documentclass{PoS}

\usepackage{bm}
\usepackage{amsfonts,amssymb,amsmath}

\newcommand{\be}{\begin{equation}}
\newcommand{\bea}{\begin{eqnarray}}
\newcommand{\ee}{\end{equation}}
\newcommand{\eea}{\end{eqnarray}}
\newcommand{\bpi}{\begin{picture}}
\newcommand{\bce}{\begin{center}}
\newcommand{\epi}{\end{picture}}
\newcommand{\ece}{\end{center}}

\newcommand{\sla}{\slash \hspace{-0.22cm}}

\title{Chiral symmetry breaking with \\ a non-enhanced ghost sector}

\ShortTitle{Chiral symmetry breaking with a non-enhanced ghost sector}

\author{\speaker{Arlene ~C. Aguilar}\\
        Federal University of ABC, CCNH, \\
Rua Santa Ad\'{e}lia 166, CEP 09210-170, Santo Andr\'{e}, Brazil.\\ 
        E-mail: \email{Arlene.Aguilar@ufabc.edu.br}}

\abstract{We study chiral symmetry breaking  using the quark gap equation supplemented with 
the infrared-finite gluon propagator and ghost dressing function obtained from large-volume  lattice simulations. One of the most 
important ingredients of this analysis is the non-Abelian quark-gluon vertex, which displays a crucial dependence on the ghost dressing function and the quark-ghost scattering amplitude. The various theoretical ingredients necessary for this construction are reviewed in detail. As a result, we obtain a dynamical quark masses of the order of 300 MeV, which is used to compute phenomenological parameters such as the pion decay constant and the quark condensate.}

\FullConference{The many faces of QCD\\
		November 2-5, 2010\\
		Gent Belgium}

\begin{document}

\section{Introduction}

One of the most crucial features of QCD is 
the dynamical generation of a quark mass and the related phenomenon of   
chiral symmetry breaking (CSB). The realization of these characteristic QCD phenomena  
is intimately connected with the non-perturbative nature of the infrared sector of the theory; therefore, 
their study in the continuum is best studied within the inherently non-perturbative formalism based on the 
Schwinger-Dyson Equations (SDE)~\cite{Aguilar:2010cn, Cornwall:1989gv}.

It is well known that the  non-linear integral equation governing 
the behavior of quark propagator (gap equation) displays a very rich structure. In particular,    
the existence or not of  non-trivial solutions for this equation depends crucially on the strength of its kernel, 
which is largely determined by the non-perturbative behavior of the  fully dressed gluon propagator 
and the quark-gluon  vertex~\cite{Aguilar:2010cn}.
Actually, the quark-gluon vertex plays 
an absolutely essential role, introducing to the gap equation a
dependence on  the ghost dressing function~\cite{Fischer:2003rp} and 
the quark-ghost scattering amplitude~\cite{Aguilar:2010cn}; as we will see, this dependence turns out to be numerically crucial 
for obtaining phenomenologically acceptable quark masses. 

In this talk, we will show how it is possible to obtain physically relevant solutions for the quark dynamical mass using the 
infrared finite ingredients obtained from the lattice, i.e., a non-vanishing gluon propagator and a  non-enhanced ghost sector.
To do that, we build a truncation scheme for the  quark  SDE,
supplemented  with three non-perturbative ingredients: (i)~the gluon propagator, (ii) the ghost
dressing function obtained  from large-volume lattice simulations~\cite{Bogolubsky:2007ud, Cucchieri:2007md}, and (iii)
the ``one-loop dressed'' approximate version of the scalar form factor of the 
quark-ghost scattering kernel~\cite{Aguilar:2010cn}. In addition,  the results for the fermion masses in the  adjoint representation 
are briefly discussed.

\section{The gap equation}

Let us start by defining some basic quantities 
that are important for the analysis of the CSB.
The quantity  that will be on the focus of our attention is the 
full quark propagator;  in Minkowski space, its inverse  
has the general form given by
\be
S^{-1}(p) = \sla{p} -m -\Sigma(p) = A(p^2)\,\sla{p} - B(p^2) \,,
\label{qpropAB}
\ee
where $m$ is the bare current quark mass, and  $\Sigma(p)$  the quark
self-energy. Notice that  the self-energy can be decomposed  in terms of a Dirac vector component, $A(p^2)$, and a scalar
component, $B(p^2)$, which allow us to define the dynamical quark mass function as being the ratio   
\mbox{${\mathcal{M}}(p^2)= B(p^2)/A(p^2)$}.

As we are interested in generating the quark mass exclusively 
through dynamical effects,  we consider  the case without explicit CSB, where the 
bare mass $m = 0$. Then, it is easy 
to see that the quark propagator will develop a dynamical 
quark mass only if the scalar component, $B(p^2)$, is different from zero~\cite{Roberts:1994dr}. 

\begin{figure}[!h]
\begin{center}
\resizebox{0.7\columnwidth}{!}
{\includegraphics{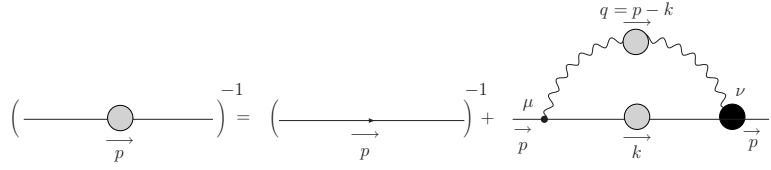}}
\end{center}
\caption{Diagrammatic representation of the quark SDE  (gap equation).}
\label{gap_eq}
\end{figure}

The diagrammatic representation of  $S^{-1}(p)$ is shown
in Fig.\ref{gap_eq}; using the same convention of  momenta flow as indicated in the figure, 
the gap equation can be written as
\begin{equation}
S^{-1}(p)= \sla{p} -C_{\rm r}g^2\int_k
\Gamma_{\mu}^{[0]}S(k)\Gamma_{\nu}(-p,k,q)\Delta^{\mu\nu}(q) \,,
\label{senergy}
\end{equation}
where $q\equiv p-k$, \mbox{$\int_{k}\equiv\mu^{2\varepsilon}(2\pi)^{-d}\int\!d^d k$}, 
with $d=4-\epsilon$ the dimension of space-time. $C_{\rm r}$  is the Casimir eigenvalue of 
the given fermion representation, where we set  
$r={\rm F}$  for the fundamental, and $r{=\rm A}$ for the  adjoint. Specifically, for the gauge 
group  $SU(3)$, we have \mbox{$C_{\rm A}=3$} and \mbox{$C_{\rm F}=4/3$}.

According to Eq.~(\ref{senergy}), the quark propagator, $S(p)$, is dynamically determined in terms
of an integral equation involving  itself, the full gluon propagator, to be denoted by $\Delta^{\mu\nu}(q)$, 
the full fermion-gluon vertex $\Gamma_{\nu}(-p,k,q)$,  and its tree level counterpart  
$\Gamma_{\mu}^{[0]} = \gamma_{\mu}$.

In the Landau gauge the  full gluon propagator, $\Delta_{\mu\nu}(q)$,  has the form  
\be 
\Delta^{\mu\nu}(q)= -i\left[  g^{\mu\nu} - \frac{q^{\mu} q^{\nu}}{q^2}\right]\Delta(q^2) \,,
\label{fprop}
\ee
where the non-perturbative behavior of the scalar factor $\Delta(q^2)$ has been 
studied in great detail in the continuum 
~\cite{Aguilar:2006gr,Dudal:2008sp}, 
 and in the lattice simulations \cite{Bogolubsky:2007ud,Cucchieri:2007md}.

\subsection{The full quark-gluon vertex}

In principle, the fully-dressed quark-gluon vertex, $\Gamma_{\nu}$, is 
determined from its own SDE, which contains a number of (unknown) 
multiparticle kernels, characteristic of the so-called ``skeleton expansion''.  
Dealing with such an equation is technically very difficult; 
therefore, the standard way to obtain information about this vertex is to use 
a gauge-technique Ansatz for it~\cite{Salam:1963sa}.  
The general idea of this method is 
to express the longitudinal part of a vertex  
in terms of the various quantities appearing in the 
fundamental Ward identity or Slavnov-Taylor identity (STI) that it satisfies.

\begin{figure}[!t]
\begin{center}
\resizebox{0.4\columnwidth}{!}
{\includegraphics{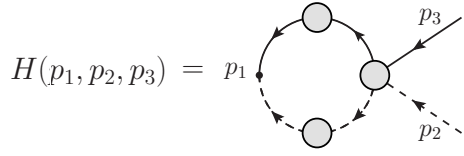}}
\end{center}
\caption{The fermion-ghost scattering kernel $H(p_1,p_2,p_3)$.}
\label{figh}
\end{figure}

In the case of the quark-gluon vertex,  the general Lorentz decomposition  
for the longitudinal part of  $\Gamma_{\nu}$  involves four 
different form factors. All these form factors contain an explicit dependence  
to both the ghost dressing function and the ``quark-ghost scattering kernel'' which, in turn,  also  has  a rich tensorial structure.
Specifically, the vertex $\Gamma_{\mu}(p_1,p_2,p_3)$
satisfies the following STI~\cite{Marciano:1977su}
\be
p_3^{\mu}\Gamma_{\mu}(p_1,p_2,p_3) = 
F(p_3)[S^{-1}(-p_1) H(p_1,p_2,p_3) - {\overline H}(p_2,p_1,p_3) S^{-1}(p_2)]\,,
\label{STI}
\ee
where \mbox{$H(p_1,p_2,p_3)$} is the quark-ghost scattering kernel
represented in the Fig.~\ref{figh}, and  $F(p_3)$ 
is the ghost dressing function which is  related to the full ghost propagator 
by \mbox{$D(p_3)= iF(p_3)/p_3^2$}. 
%

The kernel $H(p_1,p_2,p_3)$ and the 
``conjugate'' ${\overline H}(p_2,p_1,p_3)$ have the 
following Lorentz decomposition~\cite{Davydychev:2000rt} (note the change 
$p_1 \leftrightarrow p_2$ in the arguments of the latter)
\bea
H(p_1,p_2,p_3) &=& X_0 \mathbb{I}  
+X_1 \sla{p_1} +  
X_2  \sla{p_2} +
X_3 \tilde\sigma_{\mu\nu}p_1^{\mu} p_2^{\nu} \,,
\nonumber\\ 
{\overline H}(p_2,p_1,p_3) &=& 
{\overline X}_0 \mathbb{I} 
-{\overline X}_2 \sla{p_1}  
-{\overline X}_1 \sla{p_2} 
+{\overline X}_3 \tilde\sigma_{\mu\nu}p_1^{\mu} p_2^{\nu} \,,
\label{Xi}
\eea
where the form factors $X_i$ are functions of the momenta,
$X_i=X_i(p_1,p_2,p_3)$,  
and we use the notation 
${\overline X}_i (p,r,q) \equiv X_i (r,p,q)$
and $\tilde\sigma_{\mu\nu} \equiv \frac{1}{2}[\gamma_{\mu},\gamma_{\nu}]$.

The most general Lorentz decomposition for the longitudinal part of the vertex $\Gamma_{\mu}(p_1,p_2,p_3)$
can be written as~\cite{Davydychev:2000rt}
\be
\Gamma_{\mu}(p_1,p_2,p_3) = 
  L_1 \gamma_{\mu}
+ L_2 (\sla{p_1} - \sla{p_2})(p_1-p_2)_{\mu} 
+ L_3 (p_1-p_2)_{\mu} 
+ L_4 \tilde\sigma_{\mu\nu}(p_1-p_2)^{\nu} \,,
\label{Li}
\ee
where $L_i$ are the form factors, whose  dependence on the momenta 
has been suppressed, in order to keep a compact notation, {\it i.e.}, $L_i=L_i(p_1,p_2,p_3)$. Notice that the tree level expression 
for $\Gamma_{\mu}^{[0]}$  is 
recovered setting $L_1=1$ and $L_2=L_3=L_4=0$; then, 
$\Gamma_{\mu}^{[0]} = \gamma_{\mu}$.

Contracting Eq.~(\ref{Li}) with $p_3^{\mu}$, we have 
\be
p_3^{\mu}\Gamma_{\mu}(p_1,p_2,p_3) = 
(p_2^2 - p_1^2) L_3 \mathbb{I} 
+[(p_2^2 - p_1^2) L_2 - L_1]\sla{p_1}
-[(p_2^2 - p_1^2) L_2 + L_1] \sla{p_2}
- 2 L_4 \tilde\sigma_{\mu\nu}p_1^{\mu} p_2^{\nu} \,.
\label{VLi}
\ee

In addition, substituting the standard decomposition of $S^{-1}(p)$, given in Eq.~(\ref{qpropAB}), and the 
expression of $H(p_1,p_2,p_3)$ given by Eq.~(\ref{Xi}) into  Eq.~(\ref{STI}), 
we find that the rhs of Eq.~(\ref{VLi}) can be also expressed  in 
terms of the functions $A$, $B$ and $X_i$'s. Then,  it is  relatively straightforward 
to demonstrate that the $L_i$'s may be expressed as~\cite{Aguilar:2010cn}
\bea
L_1 &=& \frac{F(p_3)}{2} \bigg\{
A(p_1)[X_0 + (p_1^2- p_1\!\cdot\!p_2)X_3] 
+ A(p_2)[{\overline X}_0 +(p_2^2- p_1\!\cdot\!p_2){\overline X}_3]\bigg\} 
\nonumber\\
&+&
\frac{F(p_3)}{2} \bigg\{ B(p_1)(X_1+X_2) + B(p_2)({\overline X}_1+{\overline X}_2)\bigg\} \,;
\nonumber\\
L_2 &=& \frac{F(p_3)}{2(p_2^2 - p_1^2)} \bigg\{
A(p_1)[(p_1^2 + p_1\!\cdot\!p_2)X_3 -X_0] 
- A(p_2)[(p_2^2+p_1\!\cdot\!p_2){\overline X}_3 -{\overline X}_0]\bigg\}
\nonumber\\
&+&
\frac{F(p_3)}{2(p_2^2 - p_1^2)} \bigg\{ B(p_1)(X_2-X_1) + B(p_2)({\overline X}_1-{\overline X}_2)\bigg\} \,;
\nonumber\\
L_3 &=& - \frac{F(p_3)}{p_2^2 - p_1^2}
\bigg\{  
A(p_1) \left( p_1^2 X_1 + p_1\!\cdot\!p_2 X_2 \right)
- A(p_2) \left( p_2^2 {\overline X}_1 + p_1\!\cdot\!p_2 {\overline X}_2\right)
+ B(p_1)X_0 - B(p_2){\overline X}_0\bigg\} \,;
\nonumber\\
L_4 &=&\frac{F(p_3)}{2} \bigg\{ 
A(p_1) X_2 - A(p_2) {\overline X}_2 + B(p_1) X_3 - B(p_2){\overline X}_3 
\bigg\} \,.
\label{expLi}
\eea

It is interesting to notice that setting  in Eq.~(\ref{expLi}) 
$X_0 = {\overline X}_0=1$ and $X_i = {\overline X}_i=0$, 
for $i \geq 1$, and  $F(p_3) =1$, we obtain  
the  following expressions  
\bea
L_1 = \frac{A(p_1)+A(p_2)}{2}\,, &\qquad& L_3 = \frac{B(p_1)- B(p_2)}{p_1^2 - p_2^2} \,, 
\nonumber\\
L_2 = \frac{A(p_1)- A(p_2)}{2(p_1^2 - p_2^2)}\,,  &\qquad&  L_4 = 0 
\eea 
which give rise to the so-called Ball-Chiu (BC) vertex~\cite{Ball:1980ay}, 
\bea 
\Gamma^{\mu}_{\rm BC}(p_1,p_2,p_3)&=&\frac{A(p_1)+A(p_2)}{2}\gamma^{\mu} \nonumber \\
&+&\frac{(p_1-p_2)^{\mu}}{p_1^2-p_2^2}\left\{\left[A(p_1)-A(p_2)\right] 
\frac{\sla{p_1}-\sla{p_2}}{2}
+\left[B(p_1)-B(p_2)\right] 
\right\} \,,
\label{bcvertex}
\eea
which is widely employed in the literature  for studies of CSB~\cite{Roberts:1994dr}.

In this work, we go beyond the Abelian version of the quark-gluon vertex expressed by Eq.~(\ref{bcvertex}), and 
we consider the case where only the scalar 
component of the quark-ghost scattering kernel is non-vanishing
{\it i.e.}  \mbox{$X_0 \neq 0$} while \mbox{$X_i = {\overline X}_i=0$}, 
for $i \leq 1$. In this limit, 
the expressions of  Eq.~(\ref{expLi}) reduce to 
\bea
L_1 = F(p_3) X_0(p_3)\left[\frac{A(p_1)+A(p_2)}{2}\right] \,,  &\quad& L_3 = F(p_3) X_0(p_3)\left[\frac{B(p_1)- B(p_2)}{p_1^2 - p_2^2}\right] \,, 
\nonumber\\
L_2 = F(p_3) X_0(p_3)\left[\frac{A(p_1)- A(p_2)}{2(p_1^2 - p_2^2)}\right] \,, &\quad& L_4 = 0 \,.
\label{bci}
\eea  

According the above expression, the form factors $L_i$'s  display an 
explicit dependence on the  product $F(p_3) X_0(p_3)$ which contains
information about the infrared behavior of the ghost propagator~\cite{Aguilar:2010cn}. Therefore, the information about  
the  ghost sector enters into the  gap equation Eq.~(\ref{senergy})  through the 
quark-gluon vertex of Eq.~(\ref{Li}).

Since the transverse part of the  quark-gluon vertex is not constrained 
by the STI given by Eq.~(\ref{STI}), it is possible to add a transverse 
part  to the vertex without violating the this important identity.
The so-called  Curtis and Pennington (CP) vertex~\cite{Curtis:1990zs}, to be denoted by $\Gamma^{\mu}_{\rm CP}$, 
modifies $\Gamma^{\mu}_{\rm BC}$ by an identically conserved term  \cite{Roberts:1994dr},  such that 
\bea
\Gamma^{\mu}_{\rm CP}(p_1,p_2,p_3) = \Gamma^{\mu}_{\rm BC}(p_1,p_2,p_3) 
+ \Gamma^{\mu}_{\rm T}(p_1,p_2,p_3)\,, 
\label{full_cp1}
\eea
with 
\bea
\Gamma^{\mu}_{\rm T}(p_1,p_2,p_3) = \frac{\gamma^{\mu}(p_2^2-p_1^2) -(p_1-p_2)^{\mu}(\sla{p_1}+\sla{p_2})}{2d(p_1,p_2)} 
\left[A(p_2)-A(p_1)\right] \,,
\label{cp_tran}
\eea
where 
\bea
d(p_1,p_2) = \frac{1}{p_1^2+p_2^2}\left\{(p_2^2-p_1^2)^2 + \left[\frac{B^2(p_2)}{A^2(p_2)}+\frac{B^2(p_1)}{A^2(p_1)}\right]^2\right\}\,.
\eea

In analogy to what happens in the case of the BC vertex, the ghost 
effects due to $F(p_3)$ and the quark-ghost scattering kernel  $X_0^{[1]}(p_3)$ 
will be incorporated into the CP vertex through a simple multiplication of its tensorial 
structure by  the factor  $F(p_3)X_0^{[1]}(p_3)$. Therefore, within our 
approximation, the  ``ghost-improved'' versions of the vertex, to be denoted by 
$\overline\Gamma^{\mu}_{\rm BC}(p_1,p_2,p_3)$ and $\overline\Gamma^{\mu}_{\rm CP}(p_1,p_2,p_3)$, respectively,
can be written as
\bea
\overline\Gamma^{\mu}_{\rm BC}(p_1,p_2,p_3) &=& F(p_3) X_0^{[1]}(p_3) \Gamma^{\mu}_{\rm BC}(p_1,p_2,p_3)\,,
\nonumber\\
\overline\Gamma^{\mu}_{\rm CP}(p_1,p_2,p_3) &=&  F(p_3) X_0^{[1]}(p_3)\Gamma^{\mu}_{\rm CP}(p_1,p_2,p_3)\,. 
\label{BCCPimp}
\eea
Notice that the above equation in the limit of $X_0^{[1]}(q)=1$ reduces to the  vertex employed in Ref.~\cite{Fischer:2003rp}.
 

Substituting the vertex $\overline\Gamma^{\mu}_{\rm BC}(p_1,p_2,p_3)$
given by Eqs.~(\ref{Li}) and (\ref{bci})  into the gap equation (\ref{senergy}),
and defining \mbox{$p_1=-p$}, \mbox{$p_2=k$}, and 
\mbox{$p_3=q$},  we  arrive at the following coupled system for $A(p^2)$ and $B(p^2)$ 
\bea
A(p^2)&=& 1 + C_{r}g^2  Z_c^{-1}\,\int_{k}\,
\frac{{\cal K}_0(p-k)}{A^2(k^2)k^2+B^2(k^2)}{\cal K}_A^{\rm BC}(k,p)\,, 
\label{dirac}\\ 
B(p^2) &=& C_{r}g^2  Z_c^{-1}\int_{k}\,\frac{{\cal K}_0(p-k)}{A^2(k^2)k^2+B^2(k^2)} {\cal K}_B^{\rm BC} (k,p)\,,
\label{scalar}
\eea
where the kernel ${\cal K}_0(q)$ corresponds to the part 
that is not altered by the tensorial structure of the quark-gluon vertex, namely
\bea
{\cal K}_0(q) =\Delta(q)F(q)X_0^{[1]}(q) \,, 
\label{ker1}
\eea
while the parts that are affected, ${\cal K}_A^{\rm BC}(k,p)$ and ${\cal K}_B^{\rm BC}(k,p)$, 
are given by~\cite{Hawes:1993ef}
\bea
{\cal K}_A^{\rm BC} (k,p) &=& \frac{A(k^2)}{2p^2}[A(k^2)+A(p^2)]
\left[3p\!\cdot\!k -2h(p,k) \right] -2B(k^2)\Delta B(k^2,p^2) \frac{h(p,k)}{p^2}   \nonumber \\ 
&&   -A(k^2)\Delta A(k^2,p^2)\left[k^2 -\frac{(k\!\cdot\!p)^2 }{p^2} +2\frac{k\!\cdot\!p }{p^2}h(p,k)\right]  \,,
\\
{\cal K}_B^{\rm BC} (k,p) &=& \frac{3}{2} B(k^2)[A(k^2)+A(p^2)] + 
2\left[B(k^2)\Delta A(k^2,p^2) -A(k^2)\Delta B(k^2,p^2) \right]h(p,k) \,, \nonumber
\label{kernels}
\eea
where
\be
h(p,k) \equiv \frac{\left[k^2p^2-(k\!\cdot\!p)^2\right]}{q^2} \,,
\label{hf}
\ee
 and
\be
\Delta A(k^2,p^2) \equiv \frac{A(k^2)-A(p^2)}{ k^2-p^2} \,,\,\qquad
\Delta B(k^2,p^2) \equiv \frac{B(k^2)-B(p^2)}{ k^2-p^2} \,.
\label{delta}
\ee

Similarly, the effect of the vertex $\overline\Gamma^{\mu}_{\rm CP}(p_1,p_2,p_3)$ is 
to replace the kernels   
${\cal K}_A^{\rm BC}(k,p)$ and ${\cal K}_B^{\rm BC}(k,p)$,  
appearing in Eq.~(\ref{dirac}) and (\ref{scalar}), by ${\cal K}_A^{\rm CP}(k,p)$ and ${\cal K}_B^{\rm CP}(k,p)$, respectively, where
\bea
{\cal K}_A^{CP}(k,p) &=&  {\cal K}_A^{BC}(k,p) + \frac{3k\cdot p}{2p^2}A(k^2)\Delta A(k^2,p^2)\frac{(k^2-p^2)^2}{d(k,p)} \,, \nonumber \\ 
{\cal K}_B^{CP}(k,p) &=&  {\cal K}_B^{BC}(k,p) + \frac{3}{2}B(k^2)\Delta A(k^2,p^2)\frac{(k^2-p^2)^2}{d(k,p)} \,.
\label{kcp}
\eea

\section{The non-perturbative ingredients from the lattice}

\begin{figure}[!ht]
\begin{center}
\begin{minipage}[b]{0.45\textwidth}
\includegraphics[scale=0.35]{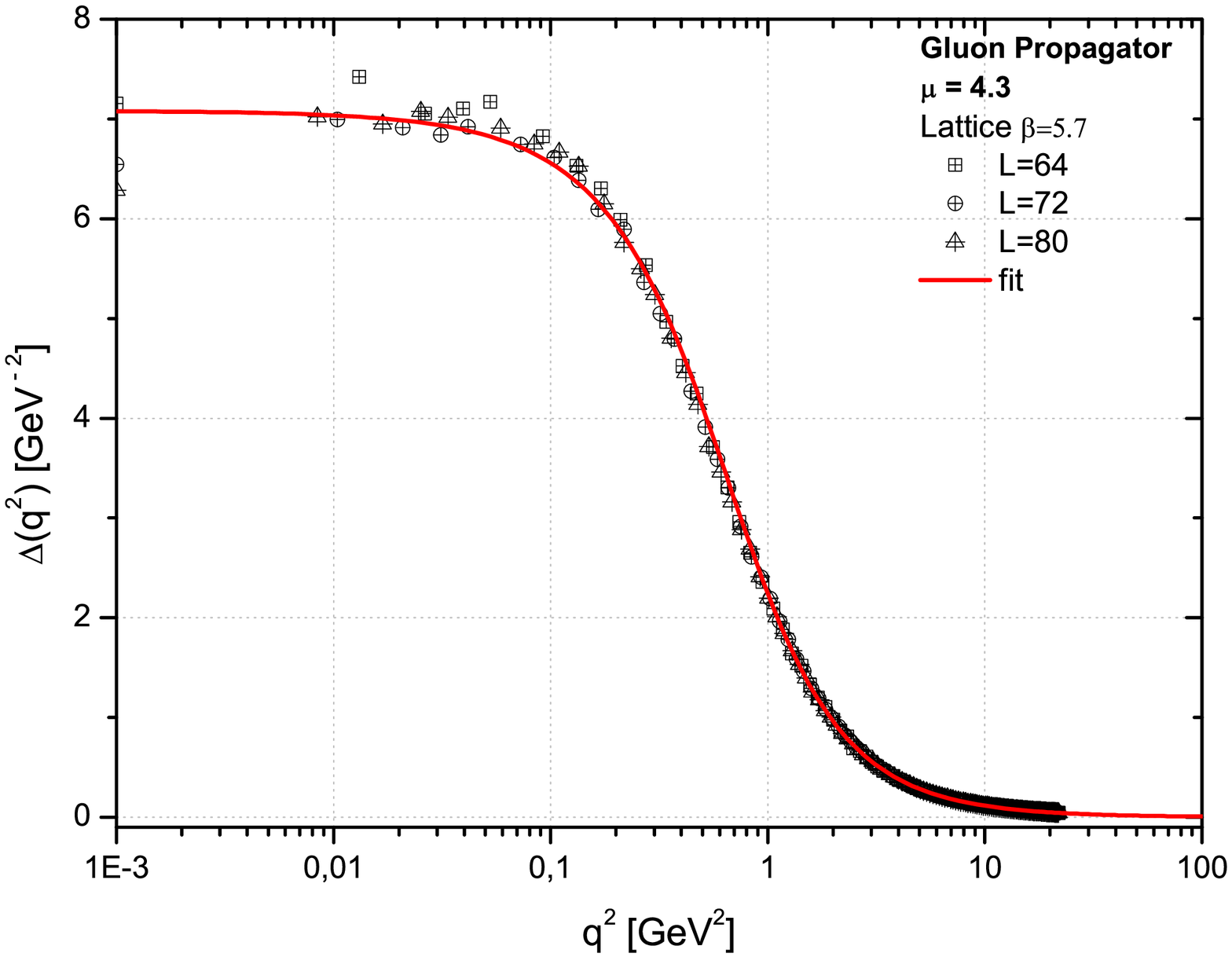}
\end{minipage}
\hspace{0.5cm}
\begin{minipage}[b]{0.50\textwidth}
\includegraphics[scale=0.35]{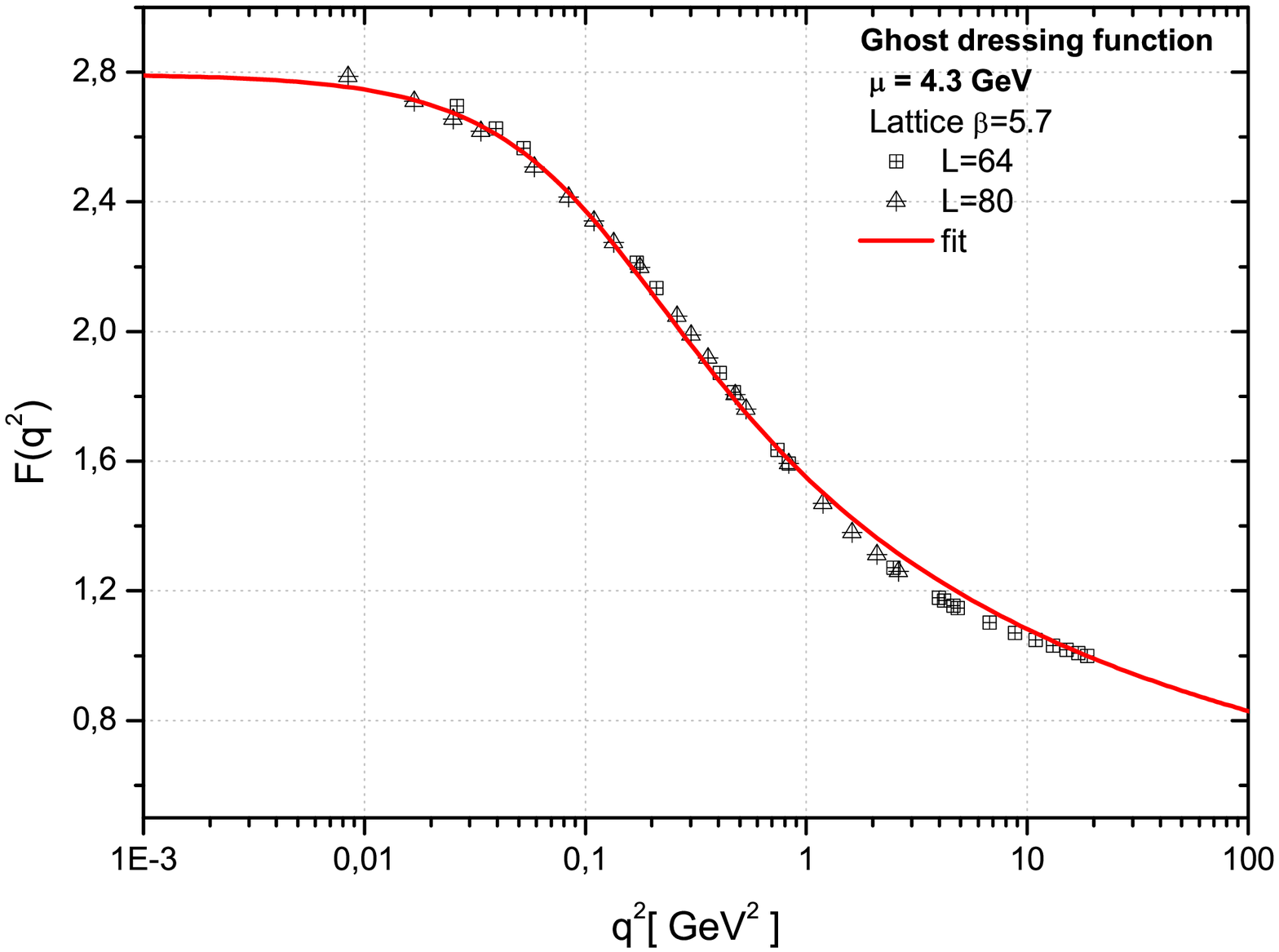}
\end{minipage}
\end{center}
\caption{Lattice results for the gluon propagator, $\Delta(q)$, and ghost dressing, $F(q)$ renormalized at \mbox{$\mu=4.3$ GeV}.}
\label{lattice}
\end{figure}

As we have seen in Eq.~{\ref{ker1}}, the gap equation given in  
Eqs.~(\ref{dirac}) and (\ref{scalar}) depends  on the nonperturbative form 
of the three basic Green's functions, namely $\Delta(q)$, $F(q)$, and $X_0(q)$.
Therefore, in order to proceed with the analysis, we use 
for $\Delta(q)$ and $F(q)$ the recent lattice data obtained by ~\cite{Bogolubsky:2007ud}, and
shown in Fig.~\ref{lattice}.

We clearly see that both lattice results for $\Delta(q)$ and $F(q)$
are infrared finite. This characteristic feature has been long  associated with a purely non-perturbative effect, 
namely the dynamical generation of an effective (momentum-dependent) 
gluon mass~\cite{Cornwall:1982zr,Cornwall:1989gv, Aguilar:2001zy}, whose presence saturates 
the gluon propagator in the IR. In addition, the  appearance of 
the gluon mass is also responsible for the infrared finiteness of the  
ghost dressing function, $F(q^2)$ \cite{Aguilar:2006gr, RodriguezQuintero:2010wy}, which is shown on the right panel
of Fig.~\ref{lattice}. 

The only ingredient missing at this point is the form factor $X_0^{[1]}(q)$. Unfortunately, 
as far we know, there are no lattice data available for $X_0(q)$. 
We will therefore proceed with our analysis by deriving an estimate for $X_0$, based on its   
``one-loop dressed'' approximation~\cite{Aguilar:2010cn}. 

\begin{figure}[!ht]
\begin{center}
\resizebox{0.2\columnwidth}{!}
{\includegraphics{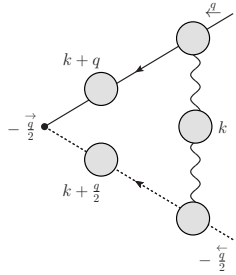}}
\end{center}
\caption{Diagrammatic representation of the quark-ghost scattering kernel, $H(-q/2,-q/2,q)$, at one-loop.}
\label{fig_sk_loop}
\end{figure}

Specifically, we will consider the scalar contribution 
of the diagram represented in Fig.~\ref{fig_sk_loop} and can be 
written as 
\be
X_0^{[1]}(p,p,q) = 1 - i\left(\frac{1}{4}\right) \frac{C_Ag^2}{2} \int_k \Delta^{\mu\nu}(k)D(k-p)\, G_{\nu}\, 
{\rm Tr}\{ S(k+q) \Gamma_{\mu}\} \,.
\label{X0LD}
\ee
where the full gluon-ghost vertex is denoted by $G_{\mu}^{ab}=\delta^{ab} G_{\mu}$. 

Using the tree-level expression for $G_{\nu} = (k-p)_{\nu}$, and the kinetic 
configuration where \mbox{$p_1 = p_2 \equiv p$}, and  $p= - q/2$,  we notice that the above equation 
simplifies considerably (for more details see \cite{Aguilar:2010cn}), 
and we arrive at 
\be
X_0^{[1]}(q) = 1 + \frac{1}{4} C_A g^2q^2 \int_k 
\left[1-\frac{(k \cdot q)^2}{k^2 q^2}\right]\Delta (k) F(k) \frac{F(k+q)}{(k+q)^4} \,.
\label{sk2t}
\ee

\begin{figure}[!h]
\begin{center}
\resizebox{0.5\columnwidth}{!}
{\includegraphics{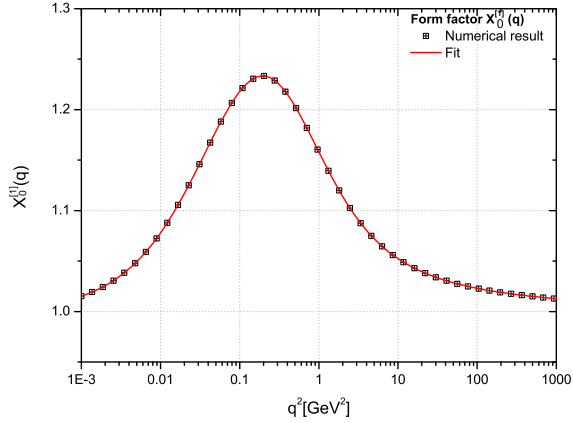}}
\end{center}
\caption{Numerical result for the form factor  $X_0^{[1]}(q)$ of  the  
 quark-ghost scattering kernel given by  Eq.~(3.2) when $\alpha(\mu^2)=0.295$.}
\label{sk}
\end{figure}

A non-perturbative estimate for $X_0^{[1]}(q)$ can be obtained by substituting into Eq.~(\ref{sk2t}) the lattice  
data for $\Delta(q)$ and $F(q)$, given in
Fig.~\ref{lattice}. The numerical result for $X_0^{[1]}(q)$ is shown in the  
Fig.~\ref{sk}.

As we can see  $X_0^{[1]}(q)$ shows a maximum  
located in the intermediate momentum region (around \mbox{$450$ MeV}),   
while in the UV and IR regions $X_0^{[1]}(q) \to 1$.
Although this peak is not very pronounced, 
it occurs in a region where the  kernel of the gap 
equation is extremely sensitive. We have noticed
that minor changes in the intermediate region result 
in considerable changes in the value of the dynamical mass
generated as discussed  in detail in Ref.~\cite{Aguilar:2010cn}).

\section{Numerical results}

\begin{figure}[!ht]
\begin{center}
\begin{minipage}[b]{0.45\textwidth}
\includegraphics[scale=0.35]{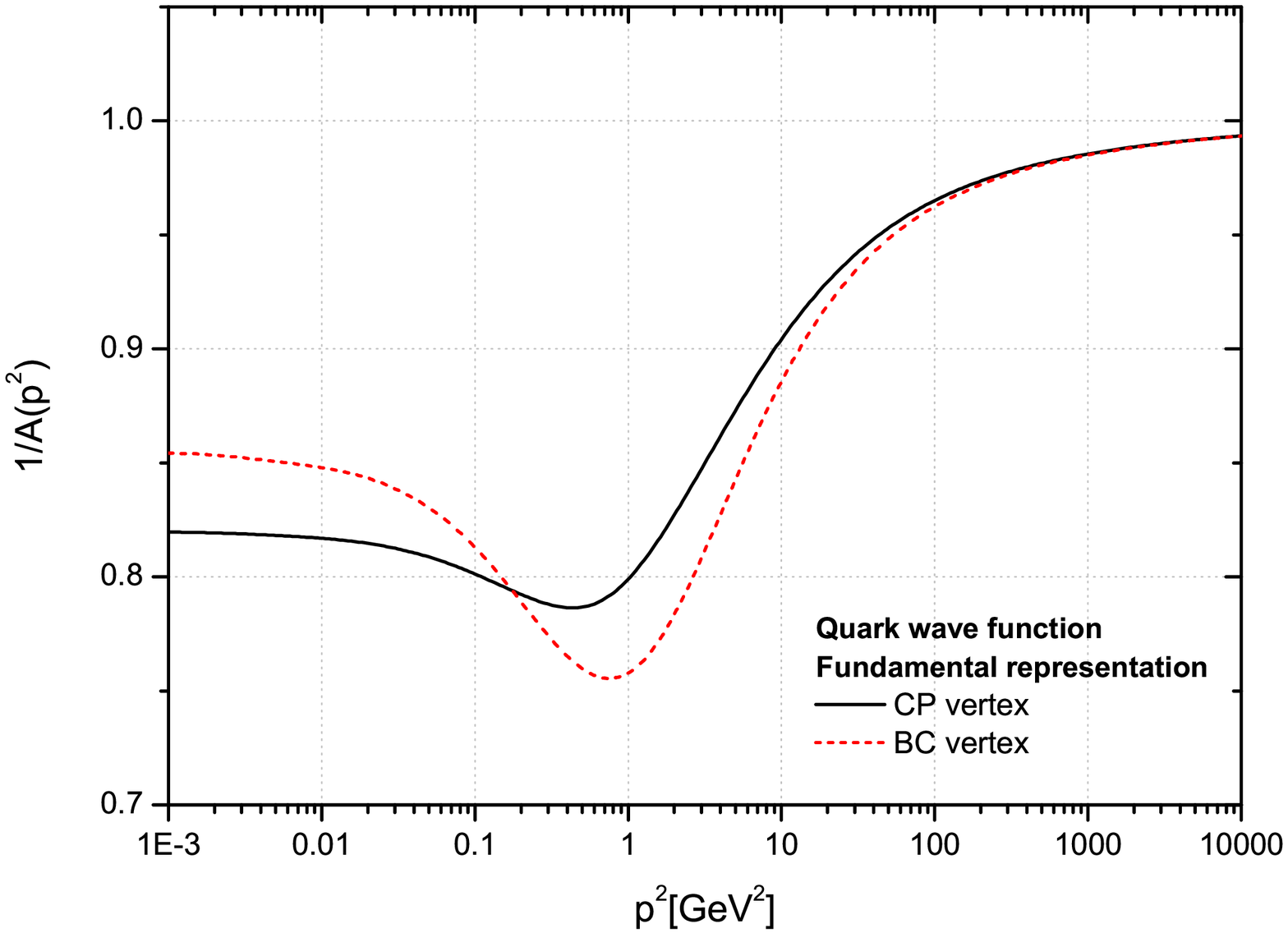}
\end{minipage}
\hspace{0.5cm}
\begin{minipage}[b]{0.50\textwidth}
\includegraphics[scale=0.35]{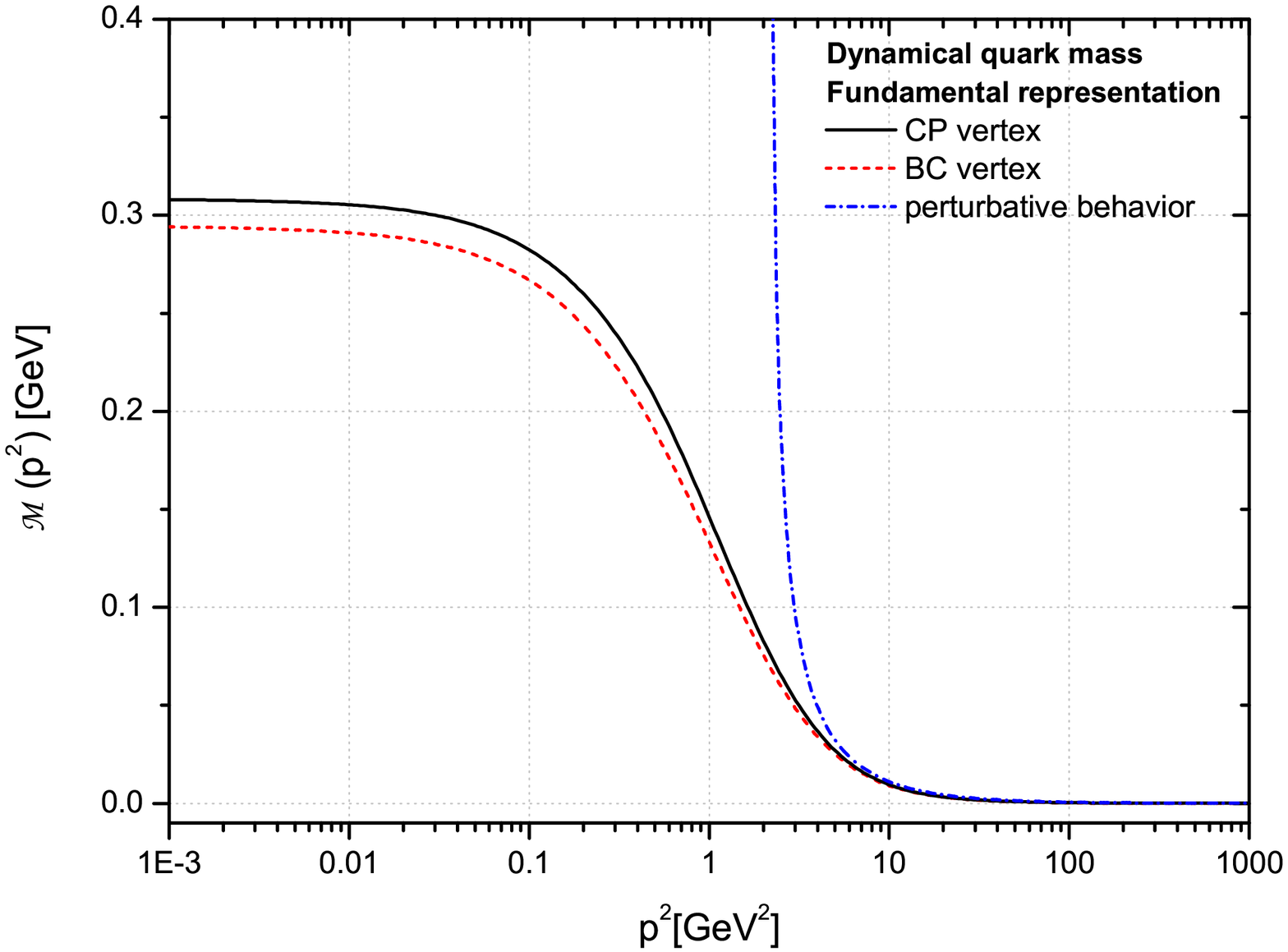}
\end{minipage}
\end{center}
\caption{{\it Left panel}: The quark wave function $A^{-1}(p^2)$ in the fundamental 
representation for both vertices  
{\it Right panel}: The  dynamical quark mass ${\mathcal M}(p^2)$.}
\label{fundamental}
\end{figure}

With all ingredients in hand, we are now in position to solve the system formed by Eqs.~(\ref{dirac}) and (\ref{scalar}). 
Substituting  $\Delta(q^2)$, $F(q^2)$, and $X_0^{[1]}(q)$ into  Eqs.~(\ref{dirac}) and (\ref{scalar}), 
with the modification \mbox{$Z_c^{-1} {\cal K}_{A,B} (k,p) \to  {\cal K}_{A,B} (k,p) F(p^2)$},
to enforce the correct renormalization group behavior of the dynamical mass (see discussion in \cite{Aguilar:2010cn}), 
we determine numerically the unknown functions $A(p^2)$ and $B(p^2)$ for the 
fundamental representation. The result
for the quark wave function $A^{-1}(p^2)$ is shown in the left panel of Fig.~\ref{fundamental}, while
the right panel shows  the dynamical quark mass ${\mathcal M}(p^2)$.

Notice that Fig.~\ref{fundamental} shows the result for both 
vertices $\overline\Gamma^{\mu}_{\rm BC}(p_1,p_2,p_3)$ (red dashed curves)  and $\overline\Gamma^{\mu}_{\rm CP}(p_1,p_2,p_3)$ (black continuous line).
One clearly  sees that ${\mathcal M}(p^2)$ 
freezes out and acquires a finite value in the IR, (\mbox{${\mathcal M}(0) = 294 $ MeV} in the case of the BC vertex,  and 
\mbox{${\mathcal M}(0) = 307 $ MeV} for the CP vertex).  Moreover,  in the UV it shows the expected 
perturbative behavior represented by the blue dashed curve.

Once the behavior of the dynamical quark mass has been determined, 
we have computed the pion decay constant and the quark condensate. For the pion decay constant  we obtained 
\mbox{$f_{\pi}= 80.6$ MeV} while for the 
condensate \mbox{$\left\langle \bar{q}q\right\rangle (1\,\mbox{GeV}^2)=\,(217 \mbox{MeV})^3$}, which are in good 
agreement with phenomenological results~\cite{Aguilar:2010cn}).

Next, we solve the system of Eqs.~(\ref{dirac}) and (\ref{scalar}) in  the adjoint representation 
{\it i.e.} \mbox{$C_{\rm r} = C_{\rm A}=3$}.  When we switch from the fundamental 
to the adjoint representation, the overall effect in the gap equation is an enhancement factor of 9/4 
due to the difference in the corresponding Casimir eigenvalues.  

\begin{figure}[!h]
\begin{center}
\begin{minipage}[b]{0.45\textwidth}
\includegraphics[scale=0.35]{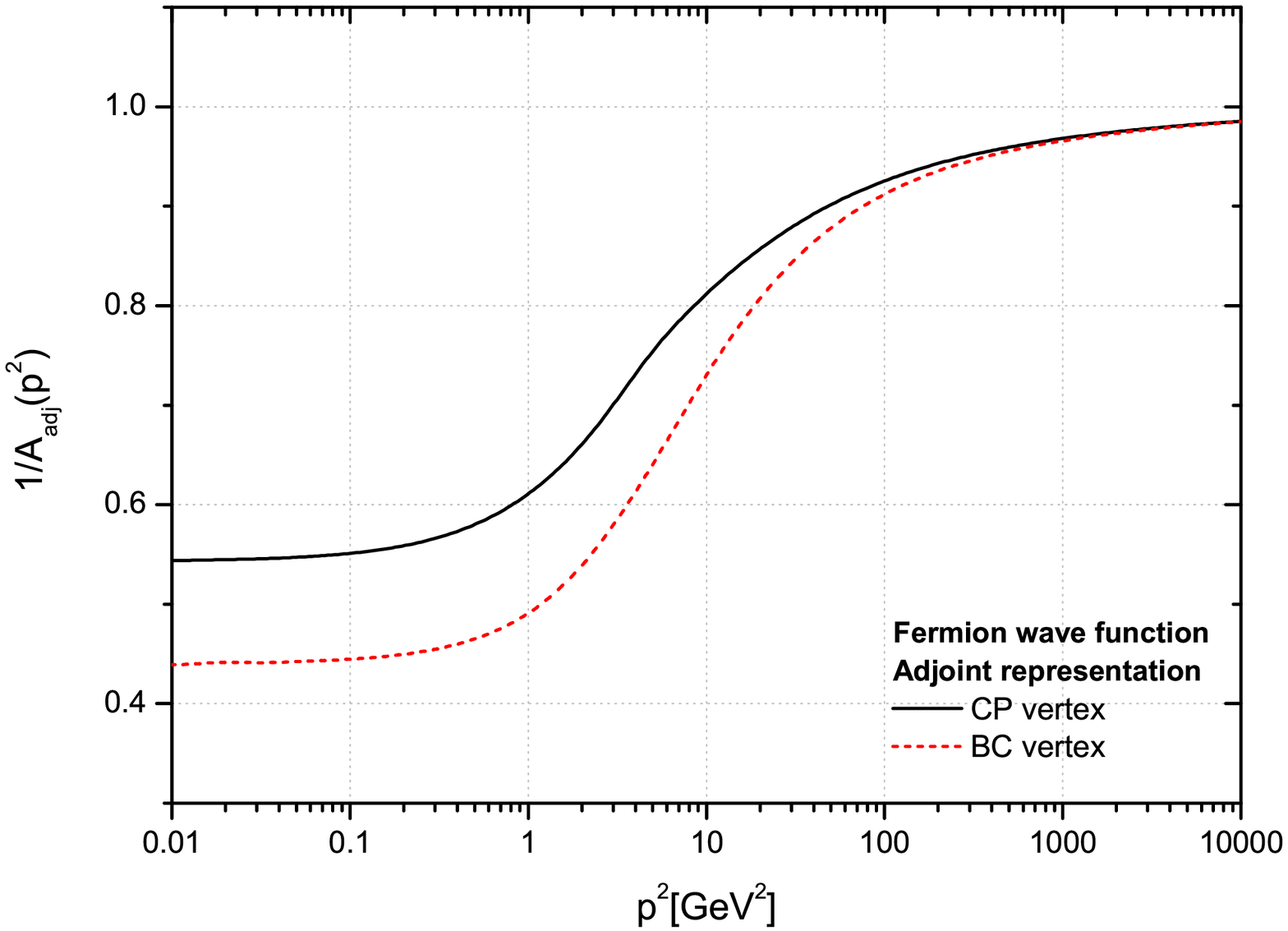}
\end{minipage}
\hspace{0.5cm}
\begin{minipage}[b]{0.50\textwidth}
\includegraphics[scale=0.35]{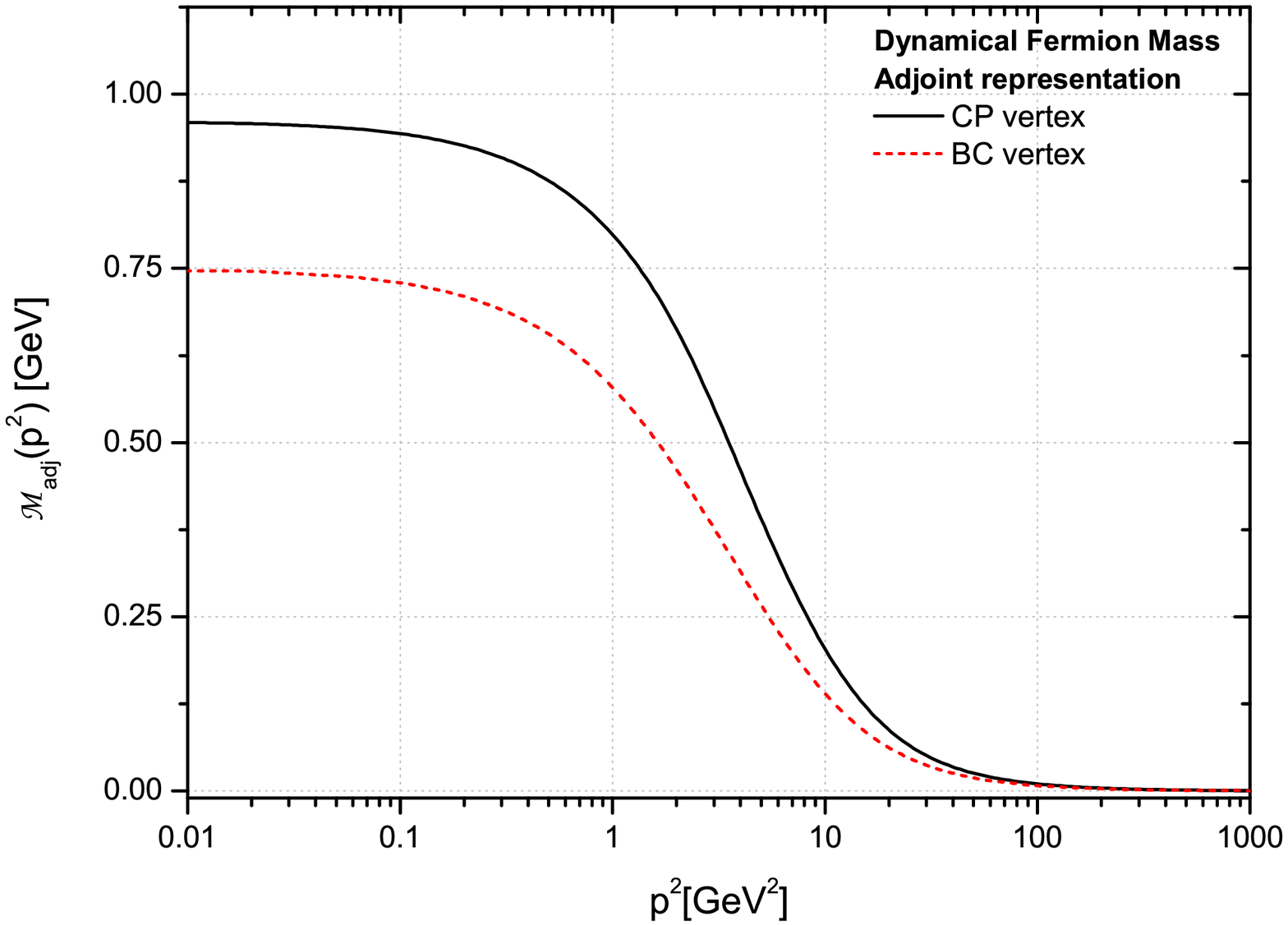}
\end{minipage}
\end{center}
\caption{{\it Left panel}: The fermion wave function $A^{-1}_{\rm adj}(p^2)$ in the 
adjoint representation. 
{\it Right panel}: The  dynamical fermion mass ${\mathcal M}_{\rm adj}(p^2)$. }
\label{adj}
\end{figure}

The numerical results for the adjoint representation 
are shown in the Fig.~\ref{adj}.  On  the left panel, we compare  
the fermion wave functions, $A^{-1}_{\rm adj}(p^2)$, when we use the modified BC (red dashed line) and 
CP vertices (black continuous line).
 
On the right panel, we show the fermion dynamical mass $\mathcal{M}_{\rm adj}(p^2)$.
We see that the infrared saturation of  $\mathcal{M}_{\rm adj}(p^2)$ occurs
for higher values compared to the  values of $\mathcal{M}(p^2)$ in the fundamental.  More
specifically, when the modified BC vertex is employed, 
one obtains \mbox{$\mathcal{M}_{\rm adj}(0)=750 $ MeV}, while for the CP vertex \mbox{$\mathcal{M}_{\rm adj}(0) = 962$ MeV}.
Clearly, due to  the  nonlinear nature of the gap equation, the results found in the adjoint representation 
can not be reproduced from the fundamental solutions through  a simple multiplication of the factor 
9/4. For example, the value for  $\mathcal{M}_{\rm adj}(0)$  are clearly higher than $9/4\mathcal{M}(0)$.

\section{Conclusions}

We have shown that the quark gap equation can give rise to 
 phenomenologically compatible results for the dynamical quark mass 
provided that (i) a complete non-Abelian quark-gluon vertex (ghost-dressing function and
quark-ghost scattering kernel) is introduced,  
and (ii) the recent lattice results for the  gluon and ghost propagators are used. 

It is important to emphasize that the incorporation of the appropriate ingredients 
in the way dictated by the underlying symmetry, as captured by the Slavnov-Taylor identity for the 
quark-gluon vertex,  
furnishes to the kernel (especially in the intermediate region of the momenta)  
the required support,
thus 
obviating the need to resort to additional (artificially induced) enhancements, 
of questionable field-theoretic origin.

\acknowledgments
I would like to thank the organizers of the Many faces of QCD for the pleasant conference. 
This research is supported by the Brazilian Funding Agency CNPq under the grants 305850/2009-1
and 453118/2010-0.


\begin{thebibliography}{99}



\bibitem{Aguilar:2010cn}
  A.~C.~Aguilar and J.~Papavassiliou,
  Phys.\ Rev.\  D {\bf 83}, 014013 (2011). 
  
  
\bibitem{Cornwall:1989gv}
  J.~M.~Cornwall and J.~Papavassiliou,
  Phys.\ Rev.\  D {\bf 40}, 3474 (1989).
  


\bibitem{Fischer:2003rp}
  C.~S.~Fischer and R.~Alkofer,
  Phys.\ Rev.\  D {\bf 67}, 094020 (2003).


\bibitem{Bogolubsky:2007ud}
  I.~L.~Bogolubsky, E.~M.~Ilgenfritz, M.~Muller-Preussker and A.~Sternbeck,
  PoS {\bf LAT2007}, 290 (2007).
  


\bibitem{Cucchieri:2007md}
  A.~Cucchieri and T.~Mendes,
  PoS {\bf LAT2007}, 297 (2007);
  O.~Oliveira and P.~J.~Silva,
  PoS {\bf QCD-TNT09}, 033 (2009).
  

\bibitem{Roberts:1994dr}
  C.~D.~Roberts and A.~G.~Williams,
  Prog.\ Part.\ Nucl.\ Phys.\  {\bf 33}, 477 (1994).
  

\bibitem{Aguilar:2006gr}
  A.~C.~Aguilar and J.~Papavassiliou,
  JHEP {\bf 0612}, 012 (2006);
  A.~C.~Aguilar, D.~Binosi and J.~Papavassiliou,
  Phys.\ Rev.\  D {\bf 78}, 025010 (2008);
  D.~Binosi and J.~Papavassiliou,
  Phys.\ Rept.\  {\bf 479}, 1 (2009).

\bibitem{Dudal:2008sp}
  D.~Dudal, J.~A.~Gracey, S.~P.~Sorella, N.~Vandersickel and H.~Verschelde,
  Phys.\ Rev.\  D {\bf 78}, 065047 (2008).


\bibitem{Marciano:1977su}
  W.~J.~Marciano and H.~Pagels,
  Phys.\ Rept.\  {\bf 36}, 137 (1978).
  


\bibitem{Salam:1963sa}
  A.~Salam,
  Phys.\ Rev.\  {\bf 130}, 1287 (1963);
  A.~Salam and R.~Delbourgo,
  Phys.\ Rev.\  {\bf 135}, B1398 (1964).

\bibitem{Davydychev:2000rt}
  A.~I.~Davydychev, P.~Osland and L.~Saks,
  Phys.\ Rev.\  D {\bf 63}, 014022 (2001).




\bibitem{Ball:1980ay}
  J.~S.~Ball and T.~W.~Chiu,
  Phys.\ Rev.\  D {\bf 22}, 2542 (1980).


\bibitem{Curtis:1990zs}
  D.~C.~Curtis and M.~R.~Pennington,
  Phys.\ Rev.\  D {\bf 42}, 4165 (1990).

\bibitem{Hawes:1993ef}
  F.~T.~Hawes, C.~D.~Roberts and A.~G.~Williams,
  Phys.\ Rev.\  D {\bf 49}, 4683 (1994).

\bibitem{Cornwall:1982zr}
J.~M.~Cornwall,
Phys.\ Rev.\ D {\bf 26}, 1453 (1982).


\bibitem{Aguilar:2001zy}
  A.~C.~Aguilar, A.~Mihara and A.~A.~Natale,
  Phys.\ Rev.\  D {\bf 65}, 054011 (2002)


\bibitem{RodriguezQuintero:2010wy}
  J.~Rodriguez-Quintero,
  JHEP {\bf 1101}, 105 (2011);
  arXiv:1012.0448 [hep-ph].


\end{thebibliography}
\end{document}